\documentclass{Interspeech}
\usepackage{comment}
\usepackage{xcolor}
\usepackage{multirow}
\usepackage{cite}
\usepackage{graphicx}
\usepackage{subcaption}
\newcommand{\sep}[1][]{\,/\,}

\interspeechcameraready

\title{ASVspoof2019 vs. ASVspoof5: Assessment and Comparison}

\author[affiliation={1}]{Avishai}{Weizman}
\author[affiliation={1}]{Yehuda}{Ben-Shimol}
\author[affiliation={2,3}]{Itshak}{Lapidot} 

\affiliation{Electrical and computer engineering}{Ben Gurion University of the Negev}{Israel}
\affiliation{Electrical engineering school}{Afeka the Academic College of Engineering in Tel Aviv}{Israel}
\affiliation{}{Avignon University}{LIA, France}

\email{wavishay@post.bgu.ac.il, benshimo@bgu.ac.il, itshakl@afeka.ac.il}

\begin{document}

\maketitle

\begin{abstract}
ASVspoof challenges are designed to advance the understanding of spoofing speech attacks and encourage the development of robust countermeasure systems. These challenges provide a standardized database for assessing and comparing spoofing-robust automatic speaker verification solutions.
The ASVspoof5 challenge introduces a shift in database conditions compared to ASVspoof2019. While ASVspoof2019 has mismatched conditions only in spoofing attacks in the evaluation set, ASVspoof5 incorporates mismatches in both bona fide and spoofed speech statistics.
This paper examines the impact of these mismatches, presenting qualitative and quantitative comparisons within and between the two databases. We show the increased difficulty for genuine and spoofed speech and demonstrate that in ASVspoof5, not only are the attacks more challenging, but the genuine speech also shifts toward spoofed speech compared to ASVspoof2019.
\end{abstract}

\section{Introduction}
\textit{Automatic speaker verification} (ASV) systems offer a cost-effective, versatile, and non-invasive biometric solution for person authentication~\cite{thomas2023broad}. Although ASV systems have achieved a level of reliability suitable for widespread adoption, concerns persist regarding their vulnerability to spoof attacks~\cite{jung2024extent,evans2013spoofing}.
Addressing these vulnerabilities necessitates robust detection solutions; however, developing and evaluating such solutions is challenging due to the significant difficulty of acquiring labeled data for speech audio attacks and authentic speaker recordings~\cite{lee2023experimental}.
The ASVspoof challenges have been introduced to address this issue and provide standardized databases and benchmarks for research in this domain~\cite{kamble2020advances}.
The ASVspoof challenges initiative was established to advance the development of spoofing detection solutions, commonly known as \textit{countermeasure} (CM) systems, to distinguish between bona fide and spoofing speech utterances~\cite{mittal2022automatic}.
It is noteworthy that the nature and severity of the attacks have evolved over the course of the ASVspoof challenges, concomitant with the increasing sophistication of the methods employed for their detection.~\cite{wang2024asvspoof,wang2025asvspoof5designcollection}.

In the context of database evaluation, it is imperative to minimize reliance on particular features to circumvent assessments that are feature-dependent. Furthermore, the evaluation process should strive for maximum independence from specific classifier structures. This approach entails the direct evaluation of databases in the time domain, with a primary focus on the waveform itself, whenever practicable.

The ASVspoof challenges were initiated in 2015~\cite{wu2014asvspoof} and have carried on in the subsequent editions in 2017~\cite{kinnunen2017asvspoof}, 2019~\cite{yamagishi2019ASVSpoof}, 2021, and most recently in 2024 with ASVspoof5~\cite{delgado2024ASVSpoof}. These challenges encompass both \textit{logical access} (LA) and \textit{physical access} (PA) attacks, with the complexity of the tasks increasing as spoofing techniques become more indistinguishable from genuine speech. In this work, our primary focus is on LA subsets of the extensively utilized database from the ASVspoof2019 challenge~\cite{yamagishi2019ASVSpoof} and the database from the most recent ASVspoof5 challenge~\cite{delgado2024ASVSpoof}.

The ASVspoof2019 LA challenge introduced advanced \textit{voice conversion} (VC) and \textit{text-to-speech} (TTS) techniques, along with their combinations~\cite{yamagishi2019ASVSpoof, wang2020asvspoof}. This challenge assumed matched conditions for spoofed speech between the training and development sets,  whereas the evaluation set had mismatched conditions, making generalization particularly challenging. The ASVspoof5 challenge posed an even greater challenge, as all three sets (training, development, and evaluation contained mismatched spoofed speech~\cite{delgado2024ASVSpoof}. Furthermore, the genuine and spoofed speech in the evaluation set was transmitted through codecs, which introduced additional mismatches for both genuine and spoofed speech.

{\let\thefootnote\relax\footnotetext{\footnotesize{\url{https://github.com/avishai111/ASVspoof2019-vs-ASVspoof5-Assessment-and-Comparison}}}

This paper presents a comparison between ASVspoof2019~\cite{yamagishi2019ASVSpoof} and ASVspoof5~\cite{wang2024asvspoof} databases. We introduce an approach to database validation and assessment, as well as a method for comparing databases. This approach involves the application of similarity measures and statistical analysis to the ASVspoof2019 and ASVspoof5 datasets. The present work draws motivation from the database assessment framework presented in~\cite{lapidot2018speech} and studies that underscore the dissimilarities between genuine and spoofed speech~\cite{lapidot2023thech,lapidot2018speech}. Moreover, we adopt and build upon several techniques in the time domain~\cite{karo2023compact,karo24_odyssey,weizmanCSCML,weizman2024tandem}. The methods demonstrated in previous studies offer straightforward, yet effective approaches for database validation. By employing \textit{PMF-based embeddings} in the time domain~\cite{karo2023compact} and utilizing dimensionality reduction as visualization tools, we facilitate the analysis of the training, development, and evaluation sets, thereby allowing a more nuanced comprehension of the complexities and dissimilarities between the datasets.

The methodologies presented here allow for the detection of matched and mismatched conditions, thereby enhancing the clarity and understanding of the database structure. Ultimately, this is expected to contribute to the development of more robust and reliable protocol designs and solutions in the future.  

The relevant literature on the use of PMF for the validation of speech databases and PMF-based time domain embeddings is examined in~\autoref{sec:related_work}. \autoref{sec:measures} presents the evaluation measures and similarity measures that are used to assess the databases that are described in~\autoref{sec:datasets}. The experimental configuration and results are presented in~\autoref{sec:Experiments_and_results}. The conclusions of the paper, detailed in~\autoref{sec:Conclusions_and_Discussion}, offer a comprehensive discourse on the findings and their implications.

\section{Related Work}
\label{sec:related_work}
In~\cite{lapidot2018speech}, a methodology is put forward for the validation of speech data sets through waveform entropy analysis. This methodology provides valuable insights that optimize database design and support the creation of more robust and accurate speech processing systems.
Furthermore, previous studies demonstrate that time-domain representations contain valuable information for various tasks, including distinguishing healthy individuals from those with early untreated Parkinson's disease, as shown in~\cite{rusz2011quantitative}. Specifically, the PMF of bona fide speech demonstrated to exhibit substantial disparities compared to that of spoofed samples, underscoring the significance of waveform distribution in anti-spoofing systems ~\cite{lapidot2019effects,lapidot2023thech,lapidot2018speech}. Building on this insight, PMF-based embeddings in the time domain were introduced in~\cite{karo2023compact,karo24_odyssey} and shown to effectively classify bona fide and spoofed speech using the ASVspoof2019 database~\cite{weizmanCSCML,weizman2024tandem}.

Furthermore, the results of the study in \cite{rohdin2024but} on the ASVspoof5 database indicate that incorporating speaker characteristics can compromise the distinction between bona fide and spoofed speech. It was observed that the variations among speakers might exceed the differences between bona fide and spoofed samples.

\section{Evaluation Measures\label{sec:measures}}
The evaluation tools use similarity measures based on class-specific PMFs or interpretable PMF-based embeddings as described in~\cite{karo2023compact}. For similarity measures based on the class-specific PMF approach, the PMF is first computed for each class in the database sets. Then, similarity measures are calculated between the different PMF classes. 
The interpretable PMF-based embeddings are generated by estimating the PMF from the amplitude distributions of bona fide and spoofed speech in the training set, processed through Gammatone and Inverse Gammatone filter banks. For each trial, similarity measures are calculated by comparing the input with class-specific PMFs as detailed in~\cite{karo2023compact}. The similarity measures employed include measures such as quadratic chi distance~\cite{pele2010quadratic}, normalized cross-correlation~\cite{briechle2001template}, Jensen-Shannon divergence~\cite{61115}, and others~\cite{pele2010quadratic,hellinger1909neue,csiszar1975divergence}. Each speech recording is represented by a 160-dimensional embedding vector, derived from 20 filters (10 Gammatone and 10 Inverse Gammatone), each contributing eight similarity measures.
It is hypothesized that, under matched conditions, the evaluation measure will have a small value. For the mismatched conditions, the measured value should be higher. In the view of the attacker, his goal is to design a spoof attack that will minimize the dissimilarity, i.e., the evaluation measure will be small.
\section{ASVspoof Databases}
\label{sec:datasets}
The following section presents the databases that will be the subject of comparison and assessment.

\subsection{ASVspoof2019 Database}
The LA attacks in the ASVspoof2019 database are partitioned into bona fide and spoofed utterances as detailed in~\autoref{tab:ASVSpoofDATA}.
The LA spoofing attacks are divided into training (Trn.), development (Dev.), and evaluation (Eval.) sets. The training and development sets share six common attack types (A01-A06), while the evaluation set contains $11$ unknown attacks (A07-A15, A17, A18), along with two attacks (A16, A19) that are based on the same TTS and VC algorithms as two attacks (A04, A06) in the training set. 
The ASVspoof2019 database contains speech data from 107 speakers, comprising 46 males and 61 females. Specifically, the training set includes data from 20 speakers (8 males, 12 females), the development set from 20 speakers (8 males, 12 females), and the evaluation set from 67 speakers (30 males, 37 females)~\cite{wang2020asvspoof}.

\subsection{ASVspoof5 Database}
The ASVspoof5 database is divided into training, development, and evaluation sets as described  in~\autoref{tab:ASVSpoofDATA}. The training set includes 8 attack types (A01-A08), the development set includes another 8 attack types (A09-A16), while the evaluation set contains 16 unknown attack types (A17-A32). Unlike previous ASVspoof challenges, ASVspoof5,  not only leverages the latest TTS and VC algorithms designed to deceive both ASV systems and CM sub-systems and for the first time, but also introduce adversarial attacks in the evaluation set as described in~\cite{wang2024asvspoof,wang2025asvspoof5designcollection}.  
Additionally, codecs, including a neural network-based codec, are applied to both bona fide and spoofed samples in the evaluation set to simulate real-world scenarios more effectively.
\begin{table}[t]
  \caption{The LA ASVspoof2019 and ASVspoof5 databases.}
  \label{tab:ASVSpoofDATA}
  \centering
\begin{tabular}{@{}lcc|cc@{}}
\toprule
\textbf{Subset} & \multicolumn{2}{c}{\textbf{ASVspoof2019}} & \multicolumn{2}{c}{\textbf{ASVspoof5}} \\ 
                & \textbf{bona fide} & \textbf{Spoof} & \textbf{bona fide} & \textbf{Spoof} \\ \midrule
\textbf{Trn.}    & 2,580  & 22,800  & 18,797  & 163,560 \\
\textbf{Dev.}    & 2,548  & 22,296  & 31,334  & 109,616 \\
\textbf{Eval.}   & 7,355  & 63,882  & 138,688 & 542,086     \\
 \bottomrule
\end{tabular}
\end{table}
The source data in the ASVspoof5 database is derived from the MLS English dataset~\cite{pratap2020mls,wang2024asvspoof,wang2025asvspoof5designcollection}, partitioned by the organizers into three disjoint subsets: A, B, and C.
The bona fide samples of the ASVspoof5 training set are from the speakers in MLS partition A, and the bona fide samples of the development and evaluation sets are from MLS partition C.
The ASVspoof5 database contains speech data from 2,659 speakers: 1,331 males and 1,328 females. 
The training set includes 400 speakers (204 males, 196 females), the development set includes 785 speakers (393 males, 392 females), and the evaluation set includes 737 speakers (367 males, 370 females).

\section{Experiments and Results}  \label{sec:Experiments_and_results}
The subsequent section delineates the experiments that were conducted and their respective results. 

The PMFs of the bona fide and spoof subsets of ASVspoof2019 and ASVspoof5 are calculated with $2^{16}$ bins as demonstrated in~\autoref{fig:ASVSpoof2019_5_combined_vertical}.
\begin{figure}[t]
    \centering
    \begin{subfigure}[b]{\linewidth}
        \centering
        \includegraphics[width=1\linewidth]{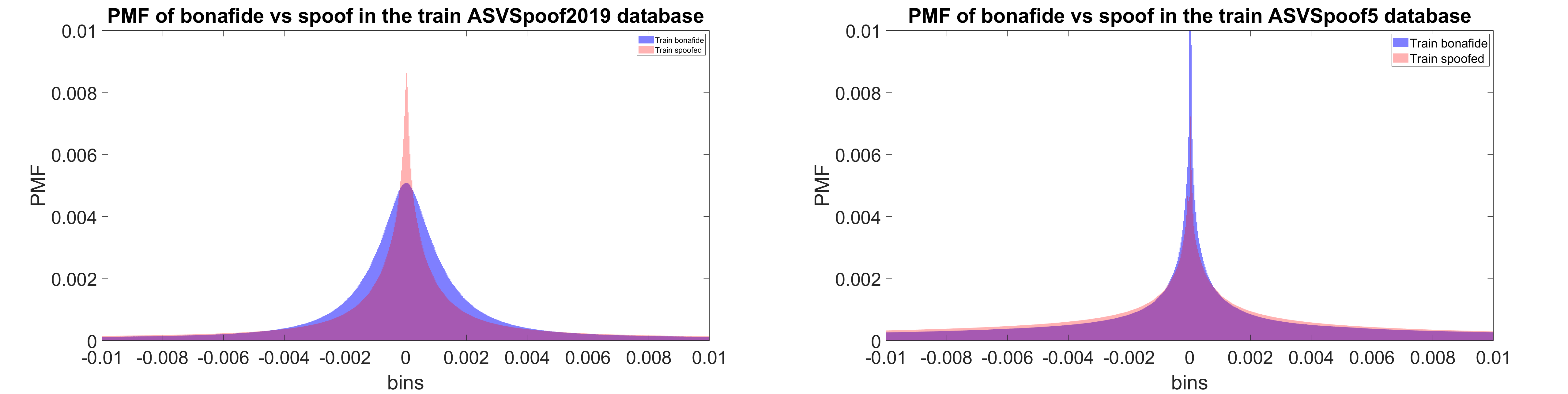}
        \caption{Training sets}
        \label{fig:train_ASVSpoof2019_5}
    \end{subfigure}
    \vspace{0.01cm}
    \begin{subfigure}[b]{\linewidth}
        \centering
        \includegraphics[width=1\linewidth]{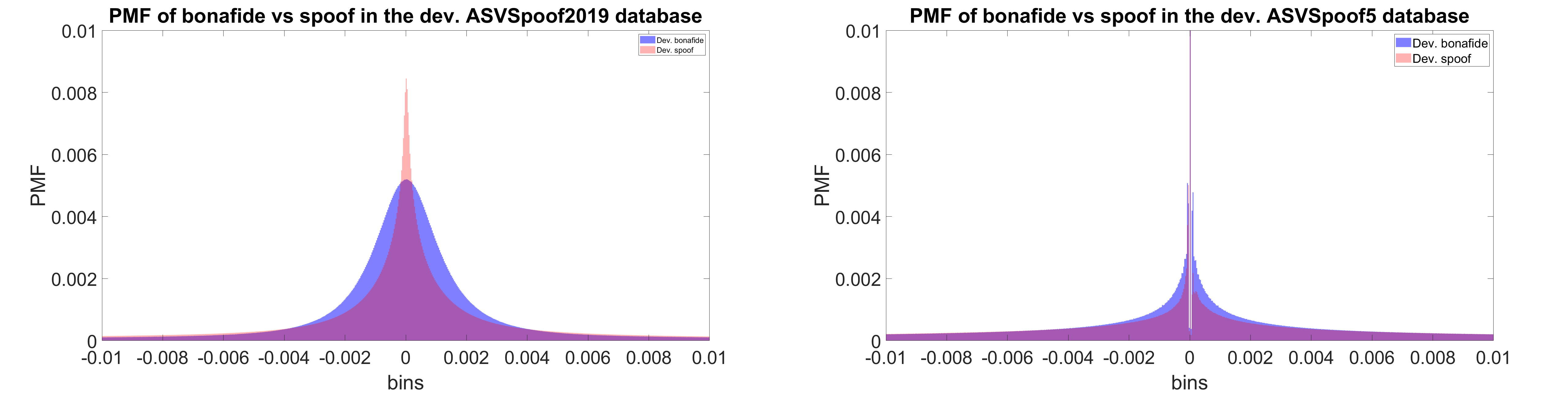}
        \caption{Development sets}
        \label{fig:dev_ASVSpoof2019_5}
    \end{subfigure}
    \vspace{0.01cm}
    \begin{subfigure}[b]{\linewidth}
        \centering
        \includegraphics[width=1\linewidth]{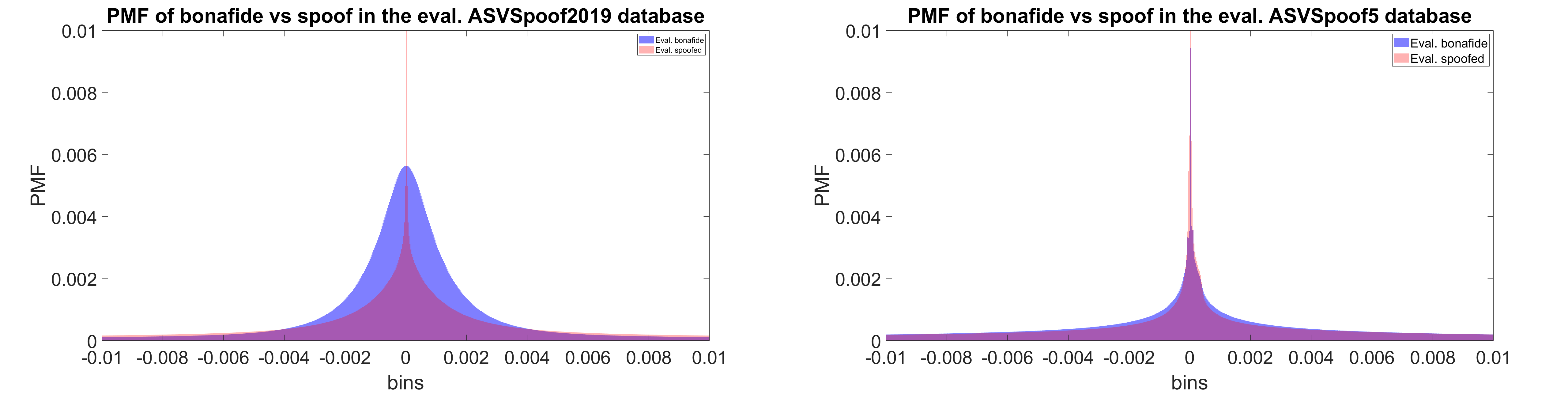}
        \caption{Evaluation sets}
        \label{fig:eval_ASVSpoof2019_5}
    \end{subfigure}
    \caption{PMF comparison for ASVspoof2019 and ASVspoof5 databases across training, development, and evaluation sets.}
    \label{fig:ASVSpoof2019_5_combined_vertical}
\end{figure}
The difference between the PMFs of bona fide and spoof utterances in~\autoref{fig:ASVSpoof2019_5_combined_vertical} is more distinct in ASVspoof2019 compared to ASVspoof5, across all sets. 
Furthermore, the PMF of bona fide speech in ASVspoof5 is significantly sharper than in ASVspoof2019. While both datasets exhibit mismatched conditions for spoof utterances in the evaluation set, ASVspoof5 demonstrates significant variation in bona fide distributions across sets, whereas ASVspoof2019 exhibits more consistent patterns (matched bona fide conditions).

Several similarity measures are calculated to compare the bona fide PMF and the spoof PMF across all subsets of the ASVspoof2019 and ASVspoof5 databases. Additionally, these similarity measures are also computed between the bona fide PMF of the ASVspoof2019 training set and the bona fide PMFs in the other subsets of ASVspoof2019 and ASVspoof5.
The measures employed include \textit{Symmetric Kullback-Leibler} (Symmetric KL)~\cite{kullback1951information}, \textit{Modified Kolmogorov-Smirnov} (Modified KS)~\cite{lapidot2020tech}, and Hellinger distance~\cite{chung1989measures}. 
The results are presented in Tables \ref{tab:Measures_datasets_each_set} and \ref{tab:Measures_datasets}.

\begin{table}[t]
    \caption{Similarity measures between the bona fide and spoof speech in all subsets of the ASVspoof2019 and ASVspoof5 databases.}
  \label{tab:Measures_datasets_each_set}
  \centering
\begin{tabular}{lcccc}
\toprule
\textbf{Database}    & \textbf{Subset}  & \textbf{Measure} & \textbf{Value}          \\ 
\midrule
ASVspoof2019             & Trn.            & Symmetric KL & 0.0650 \\
                      &                 & Modified KS            &  0.2220 \\
                      &                 & Hellinger distance                      & 0.0900 \\
\midrule
ASVspoof5             & Trn.            & Symmetric KL & 0.0537 \\
                      &                 & Modified KS            &  0.1562 \\
                      &                 & Hellinger distance                      & 0.0811 \\
\midrule                    
ASVspoof2019          & Dev.            & Symmetric KL              & 0.0967 \\
                      &                 & Modified KS             & 0.2762 \\
                      &                 & Hellinger distance                      & 0.1097 \\
\midrule
ASVspoof5             & Dev.            & Symmetric KL             & 0.0253 \\
                      &                 & Modified KS            & 0.1057 \\
                      &                 & Hellinger distance                     & 0.0561 \\
\midrule
ASVspoof2019          & Eval.           & Symmetric KL              & 0.2748 \\ 
                      &                 & Modified KS             & 0.4824 \\
                      &                 & Hellinger distance                      & 0.1845 \\
\midrule
ASVspoof5             & Eval.           & Symmetric KL             & 0.0256 \\
                      &                 & Modified KS            & 0.0642 \\
                      &                 & Hellinger distance                     & 0.0562 \\
\bottomrule
\end{tabular}
\end{table}
As illustrated in ~\autoref{tab:Measures_datasets_each_set}, the similarity measures in the training and development sets of ASVspoof2019 are comparable, likely attributable to the matched conditions. However, the measures are considerably higher on the evaluation subset, which is unexpected and may suggest that the classification task itself is not particularly challenging, with the difficulty stemming from the mismatch in conditions. Further investigation is necessary to ascertain the underlying causes. In contrast, ASVspoof5 exhibits a markedly lower similarity measure value, suggesting a closer statistical relationship between the spoof data and bona fide data. Comparing the development and evaluation subsets reveals that this relationship is further pronounced, indicating both mismatched conditions and a heightened difficulty in classification.

\begin{table}[t]
 \caption{Similarity measures based on the PMF between bona fide speech in the training set of ASVspoof2019 and the bona fide speech in all subsets of the ASVspoof2019 and ASVspoof5 databases.}
  \label{tab:Measures_datasets}
  \centering
\begin{tabular}{lcccc}
\toprule
\textbf{Database}    & \textbf{Subset}  & \textbf{Measure} & \textbf{Value}          \\ 
\midrule
ASVspoof5             & Trn.            & Symmetric KL & 0.5318 \\
                      &                 & Modified KS            & 0.5612 \\
                      &                 & Hellinger distance                      & 0.2482 \\
\midrule                    
ASVspoof2019          & Dev.            & Symmetric KL              & 0.0077 \\
                      &                 & Modified KS             & 0.0750 \\
                      &                 & Hellinger distance                      & 0.0310 \\
\midrule
ASVspoof5             & Dev.            & Symmetric KL             & 0.3655 \\
                      &                 & Modified KS            & 0.5118 \\
                      &                 & Hellinger distance                     & 0.2116 \\
\midrule
ASVspoof2019          & Eval.           & Symmetric KL              & 0.0095 \\ 
                      &                 & Modified KS             & 0.0849 \\
                     &                 & Hellinger distance                      & 0.0344 \\
\midrule
ASVspoof5             & Eval.           & Symmetric KL             & 0.4631 \\
                      &                 & Modified KS            & 0.5366 \\
                      &                 & Hellinger distance                     & 0.2150 \\
\bottomrule
\end{tabular}
\end{table}
~\autoref{tab:Measures_datasets} presents the similarity measure values between the PMF of bona fide speech in the ASVspoof2019 training set and the other PMFs of bona fide speech in the ASVspoof2019 and ASVspoof5 subsets. The similarity measure values in ASVspoof2019 remain consistent, indicating that bona fide speech in ASVspoof2019 is in a matched condition across all subsets. Conversely, the similarity measure values between the PMF of bona fide speech in the training set of ASVspoof2019 and the subsets in ASVspoof5 are considerably higher. These findings suggest a discrepancy in bona fide data collection between ASVspoof5 and ASVspoof2019.

In order to assess the impact of mismatched conditions of bona fide in ASVspoof5, the PMF-based embeddings from~\cite{karo2023compact, karo24_odyssey,weizmanCSCML, weizman2024tandem} are utilized. Subsequently, the \textit{Uniform Manifold Approximation and Projection} (UMAP) algorithm~\cite{mcinnes2018umap} is then applied to reduce the dimensionality of bona fide PMF-based embeddings in both databases to two dimensions. 
The model was trained on the ASVspoof2019 training set and subsequently applied to each subset within the ASVspoof2019 and ASVspoof5 databases.
\begin{figure}[t]
    \centering
    \includegraphics[width=1.0\linewidth]{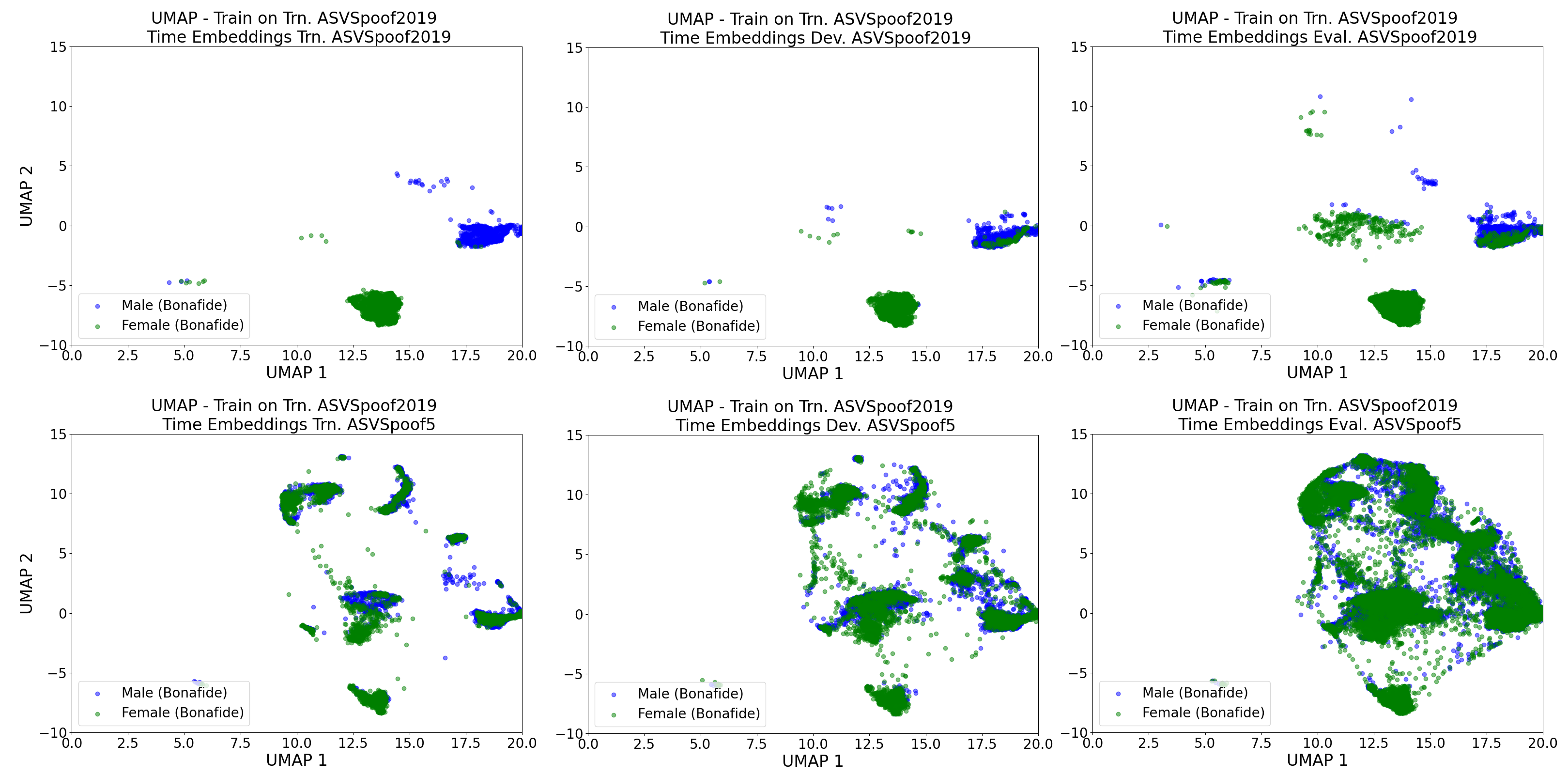}
    \caption{UMAP trained on the ASVspoof2019 training set and subsequently applied to the bona fide utterances contained within the ASVspoof2019 and ASVspoof5 databases.}
    \label{fig:time_embeddings_ASVSpoof2019_05}
\end{figure}
As illustrated in~\autoref{fig:time_embeddings_ASVSpoof2019_05}, the embeddings of males and females are presented separately. It is apparent that the bona fide PMF-based embedding from the ASVspoof2019 database demonstrates greater separation between genders and reduced variability compared to the bona fide PMF-based embedding from the ASVspoof5 database.
Additionally, it is evident that the bona fide in ASVspoof5 exhibits significant deviation from ASVspoof2019, potentially attributable to mismatched conditions.
Furthermore, the PMF-based embeddings of the ASVspoof5 evaluation set exhibit a more extensive distribution, suggesting that the bona fide training and development sets are contained within the evaluation set. This finding indicates mismatched conditions, thereby making the bona fide in the evaluation set more challenging to detect.

In the subsequent comparison, the UMAP algorithm is utilized on the PMF-based embedding derived from the utterances in the evaluation set of the ASVspoof5 database that were not processed by any codec, as depicted in~\autoref{fig:time_embeddings_codec}.
\begin{figure}[t]
    \centering
    \includegraphics[width=1.0\linewidth]{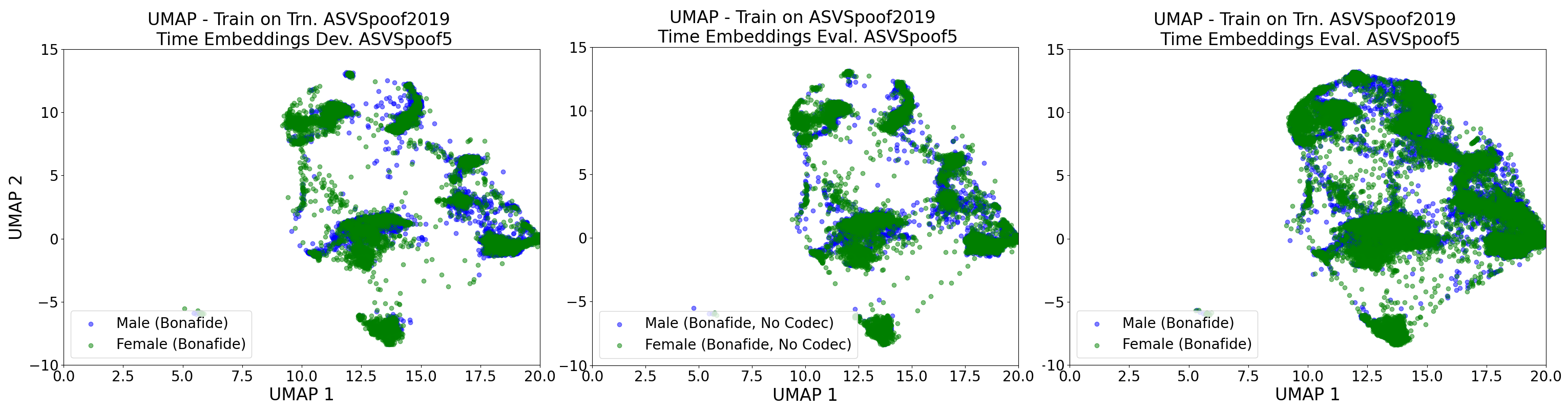}
    \caption{UMAP trained on the ASVspoof2019 training set and applied to the bona fide utterances in the Dev., Eval. (no codec), and Eval. sets of ASVspoof5.}
    \label{fig:time_embeddings_codec}
\end{figure}
The uncompressed bona fide PMF-based embeddings from the ASVspoof5 database in~\autoref{fig:time_embeddings_codec} are denser than those of compressed utterances and are similar to the plot of the development set. This observation underscores the impact of compression on the distribution's spread in the ASVspoof5 database's evaluation set.

The mismatches between the bona fide PMFs in the ASVspoof5 database introduce a new level of complexity in detecting bona fide utterances and pose an additional challenge for CM solutions.
To evaluate this impact, we implemented the system from~\cite{zhang2021one}, which was trained with \textit{one-class softmax} (OCS) loss on the training set of the ASVspoof2019 database.
The threshold was determined based on the \textit{equal error rate} (EER) of the development set of the ASVspoof2019 database. The confidence intervals were calculated using the method described in~\cite{Confidence_Intervals}, with $\alpha = 5$ indicating the confidence level, and using 1000 bootstraps. 
The misclassification rate (miss rate) of the bona fide utterances is shown in~\autoref{tab:EER_datasets}.
\begin{table}[t]
  \caption{The miss rate of the OCS CM system from~\cite{zhang2021one} on the bona fide utterances from ASVspoof2019 and ASVspoof5 databases. The threshold was determined based on the EER of the development set in the ASVspoof2019.}
  \label{tab:EER_datasets}
  \centering
\begin{tabular}{lcccc}
\toprule
\textbf{Database}      & \textbf{Subset} & \textbf{Gender} & \textbf{Miss Rate (\%)}          \\ 
\midrule
ASVspoof2019          & Dev.            & Male            & 0.46\% {\scriptsize [0.11\%, 1.76\%]} \\
                      &                 & Female          & 0.17\% {\scriptsize [0.00\%, 0.62\%]} \\
                      &                 & All             & 0.27\% {\scriptsize [0.07\%, 0.85\%]} \\ 
\midrule
ASVspoof5             & Dev.            & Male            & 58.27\%  {\scriptsize [55.21\%, 61.33\%]} \\
                      &                 & Female          & 61.95\%  {\scriptsize [59.00\%, 64.88\%]} \\
                      &                 & All             & 60.10\% {\scriptsize [58.10\%, 62.20\%]} \\ 
\midrule
ASVspoof2019          & Eval.            & Male            & 0.13\% {\scriptsize [0.00\%, 0.43\%]} \\ 
                      &                 & Female          &  0.21\% {\scriptsize [0.03\%, 0.51\%]} \\ 
                      &                 & All             &  0.19\% {\scriptsize [0.04\%, 0.39\%]} \\ 
\midrule
ASVspoof5             & Eval.            & Male            & 88.93\% {\scriptsize [88.05\%, 89.89\%]} \\
                      &                 & Female          & 89.82\% {\scriptsize [88.99\%, 90.67\%]} \\
                      &                 & All             & 89.39\% {\scriptsize [88.77\%, 90.04\%]} \\ 
\bottomrule
\end{tabular}
\end{table}
The miss rate of bona fide utterances in~\autoref{tab:EER_datasets} is significantly higher in the ASVspoof5 database. This indicates a substantial mismatch between the bona fide in the ASVspoof5 database and those in the ASVspoof2019 database, suggesting potential variations in their respective distributions and characteristics. These variations may be attributed to factors such as recording conditions, speech attributes, speaker profiles, or other attributes.

\section{Conclusions and Discussion} \label{sec:Conclusions_and_Discussion}
This study proposes a novel approach to database evaluation, assessment, and comparison. The evaluation is based on waveform PMF distributions, similarity measures, PMF-based embeddings, and UMAP visualization. In this study, the proposed methodology is employed to make a comparison between ASVspoof2019 and ASVspoof5 databases. The study introduces several approaches to assessing mismatched conditions and proposes protocols to ensure fair evaluation standards. The findings show significant differences between the two databases. ASVspoof2019 demonstrated matched conditions for bona fide samples in the training, development, and evaluation sets, whereas ASVspoof5 does not. This discrepancy helps explain the relatively high results achieved in the ASVspoof2019 challenge compared to those in the ASVspoof5 challenge. The utilization of the UMAP algorithm for the visualization of the PMF-based embeddings revealed substantial contrasts in data distribution, with the embeddings in ASVspoof2019 displaying tightly clustered structures with reduced variability, thereby facilitating more effective differentiation between bona fide and spoof utterances. In contrast, the PMF-based embeddings in ASVspoof5 exhibited a more extensive distribution.  Furthermore, an analysis of the uncompressed bona fide utterances in ASVspoof5 revealed that they exhibited more concentrated and less dispersed clusters compared to the compressed bona fide utterances. This finding indicates that compression significantly impacts the distribution of the evaluation set, thereby adding additional layers of complexity to the anti-spoofing task.
These findings underscore the importance of validating matched and mismatched conditions in databases and demonstrate how this can be achieved. The similarity measures between the bona fide and spoof PMFs of the ASVspoof5 are much smaller than those of the ASVspoof2019, thus introducing a new, and increasing level of complexity from Train. to Dev. to Eval. sets.
\section{Acknowledgements}
This work is supported by the Israel Innovation Authority under project numbers 82457 and 82458.

\bibliographystyle{IEEEtran}
\bibliography{references}

\begin{thebibliography}{10}
\providecommand{\url}[1]{#1}
\csname url@samestyle\endcsname
\providecommand{\newblock}{\relax}
\providecommand{\bibinfo}[2]{#2}
\providecommand{\BIBentrySTDinterwordspacing}{\spaceskip=0pt\relax}
\providecommand{\BIBentryALTinterwordstretchfactor}{4}
\providecommand{\BIBentryALTinterwordspacing}{\spaceskip=\fontdimen2\font plus
\BIBentryALTinterwordstretchfactor\fontdimen3\font minus \fontdimen4\font\relax}
\providecommand{\BIBforeignlanguage}[2]{{%
\expandafter\ifx\csname l@#1\endcsname\relax
\typeout{** WARNING: IEEEtran.bst: No hyphenation pattern has been}%
\typeout{** loaded for the language `#1'. Using the pattern for}%
\typeout{** the default language instead.}%
\else
\language=\csname l@#1\endcsname
\fi
#2}}
\providecommand{\BIBdecl}{\relax}
\BIBdecl

\bibitem{thomas2023broad}
P.~A. Thomas and K.~Preetha~Mathew, ``A broad review on non-intrusive active user authentication in biometrics,'' \emph{Journal of Ambient Intelligence and Humanized Computing}, vol.~14, no.~1, pp. 339--360, 2023.

\bibitem{jung2024extent}
J.-w. Jung, X.~Wang, N.~Evans, S.~Watanabe, H.-j. Shim, H.~Tak, S.~Arora, J.~Yamagishi, and J.~S. Chung, ``To what extent can asv systems naturally defend against spoofing attacks?'' \emph{arXiv preprint arXiv:2406.05339}, 2024.

\bibitem{evans2013spoofing}
N.~W. Evans, T.~Kinnunen, and J.~Yamagishi, ``Spoofing and countermeasures for automatic speaker verification,'' in \emph{INTERSPEECH 2013, 14th Annual Conference of the International Speech Communication Association}, 2013.

\bibitem{lee2023experimental}
Y.~Lee, N.~Kim, J.~Jeong, and I.-Y. Kwak, ``Experimental case study of self-supervised learning for voice spoofing detection,'' \emph{IEEE Access}, vol.~11, pp. 24\,216--24\,226, 2023.

\bibitem{kamble2020advances}
M.~R. Kamble, H.~B. Sailor, H.~A. Patil, and H.~Li, ``Advances in anti-spoofing: from the perspective of asvspoof challenges,'' \emph{APSIPA Transactions on Signal and Information Processing}, vol.~9, p.~e2, 2020.

\bibitem{mittal2022automatic}
A.~Mittal and M.~Dua, ``Automatic speaker verification systems and spoof detection techniques: review and analysis,'' \emph{International Journal of Speech Technology}, vol.~25, no.~1, pp. 105--134, 2022.

\bibitem{wang2024asvspoof}
X.~Wang, H.~Delgado, H.~Tak, J.-w. Jung, H.-j. Shim, M.~Todisco, I.~Kukanov, X.~Liu, M.~Sahidullah, T.~Kinnunen \emph{et~al.}, ``{ASVspoof} 5: Crowdsourced speech data, deepfakes, and adversarial attacks at scale,'' \emph{arXiv preprint arXiv:2408.08739}, 2024.

\bibitem{wang2025asvspoof5designcollection}
\BIBentryALTinterwordspacing
X.~Wang, H.~Delgado, H.~Tak, J.~weon Jung, H.~jin Shim, M.~Todisco, I.~Kukanov, X.~Liu, M.~Sahidullah, T.~Kinnunen, N.~Evans, K.~A. Lee, J.~Yamagishi, M.~Jeong, G.~Zhu, Y.~Zang, Y.~Zhang, S.~Maiti, F.~Lux, N.~Müller, W.~Zhang, C.~Sun, S.~Hou, S.~Lyu, S.~L. Maguer, C.~Gong, H.~Guo, L.~Chen, and V.~Singh, ``Asvspoof 5: Design, collection and validation of resources for spoofing, deepfake, and adversarial attack detection using crowdsourced speech,'' 2025. [Online]. Available: \url{https://arxiv.org/abs/2502.08857}
\BIBentrySTDinterwordspacing

\bibitem{wu2014asvspoof}
Z.~Wu, T.~Kinnunen, N.~Evans, and J.~Yamagishi, ``{ASVspoof} 2015: Automatic speaker verification spoofing and countermeasures challenge evaluation plan,'' \emph{Training}, vol.~10, no.~15, p. 3750, 2014.

\bibitem{kinnunen2017asvspoof}
T.~Kinnunen, M.~Sahidullah, H.~Delgado, M.~Todisco, N.~Evans, J.~Yamagishi, and K.~A. Lee, ``The asvspoof 2017 challenge: Assessing the limits of replay spoofing attack detection,'' in \emph{Interspeech 2017}.\hskip 1em plus 0.5em minus 0.4em\relax International Speech Communication Association, 2017, pp. 2--6.

\bibitem{yamagishi2019ASVSpoof}
J.~Yamagishi, M.~Todisco, M.~Sahidullah, H.~Delgado, X.~Wang, N.~Evans, T.~Kinnunen, K.~A. Lee, V.~Vestman, and A.~Nautsch, ``{ASVspoof} 2019: Automatic speaker verification spoofing and countermeasures challenge evaluation plan,'' \emph{ASV Spoof}, vol.~13, 2019.

\bibitem{delgado2024ASVSpoof}
H.~Delgado, N.~Evans, J.-w. Jung, T.~Kinnunen, I.~Kukanov, K.~A. Lee, X.~Liu, H.-j. Shim, M.~Sahidullah, H.~Tak \emph{et~al.}, ``{ASVspoof} 5 evaluation plan,'' 2024.

\bibitem{wang2020asvspoof}
X.~Wang, J.~Yamagishi, M.~Todisco, H.~Delgado, A.~Nautsch, N.~Evans, M.~Sahidullah, V.~Vestman, T.~Kinnunen, K.~A. Lee \emph{et~al.}, ``Asvspoof 2019: A large-scale public database of synthesized, converted and replayed speech,'' \emph{Computer Speech \& Language}, vol.~64, p. 101114, 2020.

\bibitem{lapidot2018speech}
I.~Lapidot, H.~Delgado, M.~Todisco, N.~W. Evans, and J.-F. Bonastre, ``Speech database and protocol validation using waveform entropy.'' in \emph{INTERSPEECH}, 2018, pp. 2773--2777.

\bibitem{lapidot2023thech}
I.~Lapidot and J.-F. Bonastre, ``Thech. report: Genuinization of speech waveform pmf for speaker detection spoofing and countermeasures,'' \emph{arXiv preprint arXiv:2310.05534}, 2023.

\bibitem{karo2023compact}
M.~Karo, A.~Yeredor, and I.~Lapidot, ``Compact time-domain representation for logical access spoofed audio,'' \emph{IEEE/ACM Transactions on Audio, Speech, and Language Processing}, 2023.

\bibitem{karo24_odyssey}
------, ``Meaningful embeddings for explainable countermeasures,'' in \emph{The Speaker and Language Recognition Workshop (Odyssey 2024)}, 2024, pp. 151--157.

\bibitem{weizmanCSCML}
A.~Weizman, Y.~Ben-Shimol, and I.~Lapidot, ``Spoofing-robust speaker verification based on time-domain embedding,'' in \emph{Cyber Security, Cryptology, and Machine Learning}, S.~Dolev, M.~Elhadad, M.~Kuty{\l}owski, and G.~Persiano, Eds.\hskip 1em plus 0.5em minus 0.4em\relax Cham: Springer Nature Switzerland, 2025, pp. 64--78.

\bibitem{weizman2024tandem}
------, ``Tandem spoofing-robust automatic speaker verification based on time-domain embeddings,'' \emph{arXiv preprint arXiv:2412.17133}, 2024.

\bibitem{rusz2011quantitative}
J.~Rusz, R.~Cmejla, H.~Ruzickova, and E.~Ruzicka, ``Quantitative acoustic measurements for characterization of speech and voice disorders in early untreated parkinson’s disease,'' \emph{The journal of the Acoustical Society of America}, vol. 129, no.~1, pp. 350--367, 2011.

\bibitem{lapidot2019effects}
I.~Lapidot and J.-F. Bonastre, ``Effects of waveform pmf on anti-spoofing detection,'' in \emph{Interspeech 2019}.\hskip 1em plus 0.5em minus 0.4em\relax ISCA, 2019, pp. 2853--2857.

\bibitem{rohdin2024but}
J.~Rohdin, L.~Zhang, O.~Plchot, V.~Stan{\v{e}}k, D.~Mihola, J.~Peng, T.~Stafylakis, D.~Beveraki, A.~Silnova, J.~Brukner \emph{et~al.}, ``{BUT} systems and analyses for the {ASVspoof} 5 challenge,'' \emph{arXiv preprint arXiv:2408.11152}, 2024.

\bibitem{pele2010quadratic}
O.~Pele and M.~Werman, ``{The Quadratic-Chi Histogram Distance Family},'' in \emph{Computer Vision--ECCV 2010: 11th European Conference on Computer Vision, Heraklion, Crete, Greece, September 5-11, 2010, Proceedings, Part II 11}.\hskip 1em plus 0.5em minus 0.4em\relax Springer, 2010, pp. 749--762.

\bibitem{briechle2001template}
K.~Briechle and U.~D. Hanebeck, ``Template matching using fast normalized cross correlation,'' in \emph{Optical pattern recognition XII}, vol. 4387.\hskip 1em plus 0.5em minus 0.4em\relax SPIE, 2001, pp. 95--102.

\bibitem{61115}
J.~Lin, ``Divergence measures based on the shannon entropy,'' \emph{IEEE Transactions on Information Theory}, vol.~37, no.~1, pp. 145--151, 1991.

\bibitem{hellinger1909neue}
E.~Hellinger, ``Neue begr{\"u}ndung der theorie quadratischer formen von unendlichvielen ver{\"a}nderlichen.'' \emph{Journal f{\"u}r die reine und angewandte Mathematik}, vol. 1909, no. 136, pp. 210--271, 1909.

\bibitem{csiszar1975divergence}
I.~Csisz{\'a}r, ``I-divergence geometry of probability distributions and minimization problems,'' \emph{The annals of probability}, pp. 146--158, 1975.

\bibitem{pratap2020mls}
V.~Pratap, Q.~Xu, A.~Sriram, G.~Synnaeve, and R.~Collobert, ``Mls: A large-scale multilingual dataset for speech research,'' \emph{arXiv preprint arXiv:2012.03411}, 2020.

\bibitem{kullback1951information}
S.~Kullback and R.~A. Leibler, ``On information and sufficiency,'' \emph{The annals of mathematical statistics}, vol.~22, no.~1, pp. 79--86, 1951.

\bibitem{lapidot2020tech}
I.~Lapidot, ``Tech, report: Modified kolmogorov-smirnov test,'' \emph{Journal of Biomedical Optics}, vol.~25, no.~4, pp. 1--15, 2020.

\bibitem{chung1989measures}
J.~Chung, P.~Kannappan, C.~T. Ng, and P.~Sahoo, ``Measures of distance between probability distributions,'' \emph{Journal of mathematical analysis and applications}, vol. 138, no.~1, pp. 280--292, 1989.

\bibitem{mcinnes2018umap}
L.~McInnes, J.~Healy, and J.~Melville, ``{UMAP}: Uniform manifold approximation and projection for dimension reduction,'' \emph{arXiv preprint arXiv:1802.03426}, 2018.

\bibitem{zhang2021one}
Y.~Zhang, F.~Jiang, and Z.~Duan, ``One-class learning towards synthetic voice spoofing detection,'' \emph{IEEE Signal Processing Letters}, vol.~28, pp. 937--941, 2021.

\bibitem{Confidence_Intervals}
\BIBentryALTinterwordspacing
L.~Ferrer and P.~Riera, ``Confidence intervals for evaluation in machine learning.'' [Online]. Available: \url{https://github.com/luferrer/ConfidenceIntervals}
\BIBentrySTDinterwordspacing

\end{thebibliography}

\end{document}